\documentclass[fleqn,usenatbib]{mnras}

\usepackage[T1]{fontenc}

\DeclareRobustCommand{\VAN}[3]{#2}
\let\VANthebibliography\thebibliography
\def\thebibliography{\DeclareRobustCommand{\VAN}[3]{##3}\VANthebibliography}

\usepackage{graphicx}	
\usepackage{amsmath}	
\usepackage{amssymb}	
\usepackage{hyperref}
\usepackage{physics}

\title[Stability of solar wind strahl electrons]{ Stability of superthermal strahl electrons in the solar wind}

\author[J. M. Schroeder, S. Boldyrev \& P. Astfalk]{
J. M. Schroeder,$^{1}$\thanks{E-mail: schroeder24@wisc.edu}
S. Boldyrev,$^{1,2}$
P. Astfalk,$^{3}$\thanks{Formerly at Max-Planck-Institut für Plasmaphysik, D-85748 Garching, Germany}
\\
$^{1}$Department of Physics, University of Wisconsin-Madison, Madison, WI 53706, USA\\
$^{2}$Center for Space Plasma Physics, Space Science Institute, Boulder, CO 80301, USA\\
$^{3}$2-36-6 Imado, Taito-ku, Tokyo-to 111-0024, Japan}

\date{Accepted XXX. Received YYY; in original form ZZZ}

\pubyear{2021}

\usepackage{newtxtext,newtxmath}

\begin{document}
\label{firstpage}
\pagerange{\pageref{firstpage}--\pageref{lastpage}}
\maketitle

\begin{abstract}
We present a kinetic stability analysis of the solar wind electron distribution function consisting of the Maxwellian core and the magnetic-field aligned strahl, a superthermal electron beam propagating away from the sun. We use an electron strahl distribution function obtained as a solution of a weakly collisional drift-kinetic equation, representative of a strahl affected by Coulomb collisions but unadulterated by possible broadening from turbulence. This distribution function is essentially non-Maxwellian and varies with the heliospheric distance. The stability analysis is performed with the Vlasov-Maxwell linear solver LEOPARD. We find that depending on the heliospheric distance, the core-strahl electron distribution becomes unstable with respect to sunward-propagating kinetic-Alfv\'en, magnetosonic, and whistler modes, in a broad range of propagation angles. The wavenumbers of the unstable modes are close to the ion inertial scales, and the radial distances at which the instabilities first appear are on the order of 1~AU. However, we have not detected any instabilities driven by resonant wave interactions with the superthermal strahl electrons. Instead, the observed instabilities are triggered by a relative drift between the electron and ion cores  necessary to maintain zero electric current in the solar wind frame (ion frame). Contrary to strahl distributions modeled by shifted Maxwellians, the electron strahl obtained as a solution of the kinetic equation is stable. Our results are consistent with the previous studies based on a more restricted solution for the electron strahl.  
\end{abstract}



\section{Introduction}



Observations demonstrate that the velocity distribution functions of electrons in the solar wind are essentially non-Maxwellian. They can be broken into three distinct sub-populations, an isotropic thermal core population comprising the majority of particles, a high energy nearly isotropic "halo" surrounding the core, and a tenuous field-aligned beam of particles propagating away from the sun called the "strahl" \cite[e.g.,][]{feldman75,pilipp87,salem03,maksimovic05,stverak08,pierrard2016,wilson18}.  The strahl population may carry significant outward heat flux as the solar wind expands radially outwards.  Since the net current of the system is negligible, the presence of the strahl sub-population requires a sunward drift of core electrons relative to the proton rest frame. Such an anisotropic electron distribution is expected to lead to a class of instabilities that may generate magnetosonic, Alfv\'en,  kinetic-Alfv\'en, and whistler waves. These waves are of interest because they may lead to turbulence that could, in turn, pitch-angle scatter the energetic electrons and thus regulate the electron heat flux. In particular, attention has been attracted to the whistler modes that can directly resonate with and scatter the strahl electrons  \cite[e.g.,][]{forslund70,Gary1975,Gary1994,Vocks_2005,saitogary07,garysaito07,pagel07,pierrard2011,lacombe14,Seough_2015,kajdic16,stansby16,tang2018,Verscharen_2019,boldyrev19,Lopez_2020}. However, many of the previous studies on strahl-related instabilities modeled particle distributions with Maxwellian (anisotropic, shifted Maxwellian) sub-populations. While such studies provide pivotal insight to electron instabilities in the solar wind, kinetic models generally allow for more accurate first principle studies of collisionless plasma dynamics. 

Possibly the most straightforward theoretical description of non-Maxwellian strahls is provided by kinetic exospheric models. Such models are based on the simplifying assumption that plasma close to the sun is collisional within some critical distance $r_0$ ($\sim 5-10$ solar radii), but becomes collisionless at radial distances greater than $r_0$. (In reality, this is of course an approximation since the collisionality decreases gradually in the scale of a few solar radii.) When collisions are negligible, the energy and magnetic moment of the electrons are conserved as they stream along the spatially expanding magnetic field lines \cite[][]{hollweg70,jockers70,lemaire73,scudderolbert79,maksimovic97,lie-svendsen1997,meyer-vernet1998,pierrard1999,lie-svendsen2000,scudder13,horaites18a,horaites18b,horaites2019}. These models demonstrate that the free-streaming fast electrons lead to the generation of the ambipolar electric field, so that the electrons retained by the ambipolar potential lead to the electron core formation, while those escaping it form the electron strahl. 

Such models may also be formulated to include the effects of weak Coulomb collisions and electron interactions with background turbulence. The inclusion of collisional pitch-angle scattering allows for the explanation of the strahl structure, strahl-width scaling with the electron energy and density in some observations, and also of the scaling of the electron core temperature with the heliospheric distance \cite[e.g.,][]{landi01,landi12,horaites18a,horaites18b,horaites2019,boldyrev20,bercic21}. Inclusion of strahl scattering by whistler turbulence allows one, in turn, to explain the effects of anomalously strong (that is, stronger than that predicted by Coulomb collisions) strahl broadening with the heliospheric distance and with the electron energy  \cite[e.g.,][]{hammond96,pierrard2011,anderson12,graham17,tang2018,tang20,boldyrev19,bercic19,Verscharen_2019,Lopez_2019,Micera_2020}. The possibilities have also been explored that the energetic halo component may be formed due to strong scattering and isotropization of the strahl electrons by the mechanisms mentioned above \cite[][]{Stverak2009}, or may consist of the electron population scattered from the strahl and trapped by the magnetic field at larger heliospheric distances \cite[][]{horaites2019}.

A crucial question related to the kinetic models of strahl formation deals with the stability of a plasma with such an anisotropic, beam-like component of the electron distribution. Indeed, if the strahl distribution is inherently unstable with respect to some plasma modes, then the solutions provided by the kinetic models are non-realizable as the electron strahls can be effectively destroyed (scattered) by resonant interactions with the excited waves. The unstable waves should satisfy the resonance condition, $\omega-k_\|v_\|=n\Omega_e$, where $\omega$ is the (positive) wave frequency, $\Omega_e$ is the (positive) electron cyclotron frequency, $v_\|$ is the electron velocity along the direction of the magnetic field, and $k_\|$ is the wave number along the direction of the magnetic field (we assume, without loss of generality, that the magnetic field is directed away from the sun). Particles of an antisunward moving strahl ($v_\|>0$) can interact with the whistlers through the cyclotron resonance ($n=1$) and the so-called anomalous cyclotron resonance ($n=-1$). Since for the whistlers $\omega<\Omega_e$, the cyclotron resonance is possible when $k_\|v_\|<0$, that is, the whistlers should propagate toward the sun. The anomalous cyclotron resonance is possible if $k_\|v_\|>0$. An illuminating discussion of these cases can be found  in \citet{Verscharen_2019}, where it is argued that in the former case, the whistler-related strahl instability is impossible, while in the latter the answer depends on the parameters of the electron distribution.  

In \citet{horaites18a,horaites18b,horaites2019}, the strahl distribution function was derived based on the drift-kinetic equation for the electrons with weak Coulomb collisions. The drift-kinetic equation describes the distribution function averaged over fast period of particle gyromotion in the limit when the gyroradius is much smaller than the typical scales of magnetic field variation, a situation well satisfied in the solar wind \cite[e.g.,][]{Kulsrud2005,held01,held03,smith12}.  It provides a physically realistic description of the electron strahl that deviates significantly from a shifted Maxwellian.  
The stability analysis performed in \citet{horaites18b} led to the main conclusion that the strahl electron distribution was stable, that is, it did not lead to the excitation of whistler waves resonating with the strahl. Rather, two instabilities related to low-frequency oblique magnetosonic and kinetic-Alfv\'en waves were detected. The model by \citet{horaites18b}, however, used two important approximations. First, the strahl distribution function was derived from the drift-kinetic equation at high energies but then smoothly matched at lower energies with the core electron distribution function to mimic the observations. Second, it addressed only the heliospheric distance on the order of 1 AU. A recent development of the model \cite[][]{boldyrev19,boldyrev20} allows one to derive the strahl component in a broader range of energies above the core thermal energy, and for the distances all the way down to the collisional region. 
    
In this paper, we numerically study the stability of the electron distribution function consisting of the core and strahl components, as a function of heliospheric distance. We address principal questions of whether the result of \citet{horaites18b} about the absence of strahl-resonating instabilities holds for the complete kinetic solution obtained in \citet{boldyrev19}, whether new instabilities become possible, and whether the instability thresholds depend on the heliospheric distance. Similarly to \citet{horaites18b}, we use the LEOPARD Maxwell-Vlasov solver \cite[][]{leopard} to perform a linear stability analysis for varying radial distances \footnote{We note that our focus in this study is on low-frequency modes, which are well captured by LEOPARD. We do not focus on high-frequency electrostatic modes, that as discussed in \citet{Verscharen_2019}, may not contribute to self-induced strahl scattering.}. We obtain the following results. 

First, we find that at no heliospheric distance does the strahl distribution become unstable to cyclotron resonances. This confirms and reinforces the previous result by \citet{horaites18b}. 
Second, we find that depending on the heliospheric distance, the electron velocity distribution is prone to a new quasi-parallel whistler instability, in addition to oblique fast magnetosonic modes and  kinetic Alv\'en instabilities previously also observed by \citet{horaites18b}. However, this new instability as well as the old ones found by \citet{horaites18b} are not due to strahl resonances, rather, they satisfy a Landau-Cherenkov resonance condition $\omega\approx k_\| v_{d}$, where $v_{d}$ is the drift velocity between the electron and ion velocity distribution function cores.  Such a shift in the core velocities is a general consequence of the presence of the energetic electron strahl, which ensures that the electric current is zero in the ion frame. All the unstable waves propagate in the direction of the electron-core drift, that is, in the sunward direction. An analogous whistler instability has been previously reported by \citet{Vasko_2020} based on an electron distribution model different from ours; see also a comprehensive analysis of instabilities caused by shifted Maxwellian electron distribution functions by \citet{Lopez_2020}.

Third, the obtained instabilities have thresholds that depend on heliospheric distance. For our (somewhat arbitrary but representative of the solar wind) plasma parameters, we found that all such critical distances are comparable to $1$~AU. The wavelengths of the unstable modes are found to be comparable to the ion inertial scale $d_i$, potentially making the obtained instabilities effective sources of kinetic-scale turbulence at the corresponding heliospheric distances (in agreement with previous findings by \citet{horaites18b}). The origin and structure of kinetic-scale plasma turbulence and in particular its role in solar-wind plasma heating and particle acceleration are not fully understood questions. Such questions have been addressed in many phenomenological, numerical, and observational studies  \citep[see, e.g.,][]{howes08a,howes11a,schekochihin09,alexandrova09,kiyani09a,chen10b,chen12a,chen14b,chen2020,sahraoui13a,boldyrev_etal2015,chen2016,bale16,bale2019,franci2018,Groselj2018,phan2018,passot18,kasper2019,roytershteyn19,pyakurel2019,stawarz2019,boldyrev_loureiro2019,vega20,milanese2020,Vasko_2020}. Our analysis suggests that the instabilities  exist in a broad range of angles with respect to the background magnetic field, so the generated kinetic-scale turbulence may not be restricted to parallel or oblique angles of propagation.  

\section{Electron velocity distribution}
\label{sect:2}
For our analysis, we assume a Maxwellian for the ion velocity distribution, while we represent the electron distribution function as the sum of the Maxwellian core and the strahl, $f(v_\perp, v_\|; r)=f_c(v_\perp, v_\|; r)+f_s(v_\perp, v_\|; r)$. Here,  $v_\perp$ and $v_\|$ are the velocity components perpendicular and parallel to the background magnetic field and $r$ is the heliospheric distance. Observationally, the temperature of the electron core declines with the heliospheric distance according to a power law, $T_e(r)\propto r^{-0.3}\dots r^{-0.7}$, which varies depending on whether fast or slow solar wind is considered \cite[e.g.,][]{stverak15}.  This is broadly consistent with available analytic \cite[][]{boldyrev20}  and numerical \cite[][]{bercic21} solutions of kinetic exospheric models. For our analysis, we chose $T_e(r)\propto r^{-0.5}$, although the precise value of the scaling exponent is not crucial for our conclusions. Our results do not qualitatively change for other choices of the temperature scaling. 

\begin{table}
	\caption{Radial scaling for electron and ion temperatures, density, and magnetic field strength adopted in our modeling. Ion and electron temperatures at 1 AU are taken from \citet{wilson18}. {The plasma density is assumed to be $4 $~cm$^{-3}$ at 1~AU.} {Here, $n_e(r)=n_c(r)+n_s(r)\approx n_c(r)$, since the fraction of the strahl electrons is small}. The magnetic field scaling comes from the Parker spiral model, assuming that the magnetic field lines are at $45^\circ$ to the radial direction at $r=1$~AU.}
	\centering
	\begin{tabular}{ |l| } 
		\hline\hline
		$T_e(r) = 12.21eV\ \left (\dfrac{r}{1AU}\right)^{-0.5}$\\
		\hline
	    $T_i(r) = 12.7eV\ \left (\dfrac{r}{1AU}\right)^{-0.7}$\\
	    \hline
	    $n_i(r) = n_e(r) = 4cm^{-3}\ \left (\dfrac{r}{1AU}\right)^{-2}$\\
	    \hline
	    $B(r) = B_0  \left ( \dfrac{r_0}{r} \right )^2 \sqrt{1+\left ( \dfrac{r}{1 AU} \right )^2 }$\\
		\hline\hline
	\end{tabular}
		
	\label{tab:SWscalings}
\end{table}

We emphasize here that we do not use a Maxwellian representation for the electron strahl distribution $f_s$, rather its form is derived from a first principles kinetic equation with weak Coulomb collisions. In this respect, we may call such an electron distribution a more realistic model for a {\it collisionless} plasma as compared to a shifted Maxwellian. We note that predictions of kinetic exospheric models qualitatively agree with some solar wind observations in the inner heliosphere; for instance, the relative number of particles in the strahl, the scaling of strahl angular width with the electron energy and density, and the electron temperature scaling with heliospheric distance \citep[e.g.,][]{horaites2019,boldyrev19,boldyrev20,bercic21}. We also note that our model does not include background turbulence and associated anomalous pitch-angle scattering of strahl electrons as sometimes seen in observations. This is logical, since our goal is to study whether the electron strahl predicted by a kinetic model is inherently unstable to such turbulent fluctuations in the first place. 

We approximate the background magnetic field structure by a Parker spiral, and make an assumption that the electron and ion distributions are Maxwellian in the collisional region, $r\approx r_0$, but collisions become weak at $r>r_0$. The variations of the model parameters with the radial distance are summarized in Table~\ref{tab:SWscalings}. They are fit to typical solar wind radial scaling for ion and electron temperatures \citep[][]{wilson18,stverak15}. For densities  and magnetic field strength we use scaling laws from the Parker spiral model fit to typical 1~AU values. We emphasize that although these choices do not represent the entire range of variability in the solar wind, they represent reasonable solar wind conditions \cite[e.g.,][]{cranmer09,bale16,roytershteyn19}. Small adjustments to these choices give qualitatively similar results.

The strahl distribution function has been obtained in \cite[][]{boldyrev19} as a solution of the drift-kinetic equation \cite[e.g.,][]{Kulsrud2005} with weak Coulomb pitch-angle scattering.  This function has the form:
\begin{equation}
    \begin{aligned}
    f_s(v,\theta ;r) = A_0 F_0 \dfrac{\lambda_0}{R(r)} & \left [ \dfrac{\Delta E + e \phi_\infty}{e \phi_\infty} \right ] \dfrac{\Delta E}{T_0} \text{exp} \left ( -\dfrac{\Delta E}{T_0} \right) \times \\
    &\times \text{exp} \left ( -\dfrac{{\cal E} \Delta E \text{sin}^2 \theta}{T_0^2} \dfrac{\lambda_0}{R(r)}\dfrac{B_0}{B(r)} \right),
    \end{aligned}
\label{eqn:strahl}
\end{equation}
where
\begin{equation}
   R(r) \approx r \left [ 1- 2 \left( \dfrac{T_e(r)}{\Delta E} \right ) + 2 \left( \dfrac{T_e(r)}{\Delta E} \right )^2 \text{log} \left( \dfrac{T_e(r)}{\Delta E}+1 \right )  \right ]
   \label{eqn:R(r)},
\end{equation}
and
\begin{equation}
   A_0 = n_0 \left ( \frac{m_e}{2\pi T_0} \right)^{3/2},\ \ \ \ \ \ \ F_0 = \frac{e \phi_\infty}{T_0} \text{exp} \left( - \frac{e \phi_\infty}{T_0} \right)\approx \sqrt{\frac{m_e}{m_i}}. 
   \label{eqn:Ao;Fo}
\end{equation}
Here, the velocity dependence is contained in $\Delta E = m_ev^2/2 - T_e(r)$ and ${\cal E}=m_ev^2/2$, the pitch angle $\theta$ is the angle between the local magnetic field and velocity vector, and the parameters of the plasma at $r_0$ are given in Table~\ref{tab:strahlparams}. These represent a fiducial set of parameters that illustrate the  properties of our model.  By definition, Equation~(\ref{eqn:strahl}) is valid only for $v_\parallel>0$. To avoid discontinuities in the electron distribution we use a sharp hyperbolic tangent cutoff at $v_\parallel=0$, although the particular form of the cutoff is not essential, since the distribution function is dominated by the Maxwellian core at small $v_\|$. {Indeed, the fraction of electrons in the strahl is relatively small; as was demonstrated in \cite[][]{boldyrev19,boldyrev20},  Eq.~(\ref{eqn:strahl}) leads to an estimate $n_s/n_e\propto \sqrt{m_e/m_i}$ .} At 1 AU, the resulting strahl fraction is about 5\% of the total electron population, which is consistent with observations \cite[e.g.,][]{maksimovic05,Stverak2009,graham17}.
\begin{table}
	\caption{The plasma parameters for the strahl model: $\lambda_0$ is the mean free path, $T_0$ is the electron temperature, $n_0$ is the density, and $B_0$ is the magnetic field strength at $r=r_0$.  $\Lambda$ is the Coulomb logarithm, and $\phi_\infty$ is the ambipolar potential developed in a plasma \protect\cite[][]{boldyrev20}. Values for $T_0$ and $n_0$ follow from evaluating functions in Table \ref{tab:SWscalings} at distance $r_0$. The corresponding plasma beta parameters vary with heliospheric distance in the ranges $\beta_e\sim 0.1\dots 0.3$ and $\beta_i\sim 0.2\dots0.4$.}
	\centering
	\begin{tabular}{ |c|c|c|c|c|c|c| }
		\hline
		$r_0$ & $T_0$ &$n_0$ &$B_0$ & $\lambda_0$ &  $\Lambda$ & $e \phi_\infty /
	    T_0$\\
		\hline\hline
		$5 R_\odot$  & 80 eV & $7e3\ \text{cm}^{-3}$ & 0.1042 G & $1.075 R_\odot$  & 20 & 4\\
		\hline
	\end{tabular}
	\label{tab:strahlparams}
\end{table}

\begin{figure}
    \includegraphics[width=\columnwidth]{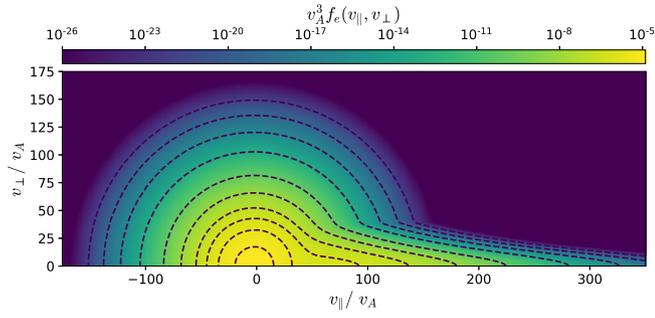}
    \caption{Full electron distribution at 1 AU normalized to unity. The distribution is comprised of an isotropic Maxwellian core and the analytical strahl model Equation \ref{eqn:strahl}. Note that the $v_\perp$ width of the strahl decreases as $v_\|$ increases.}
    \label{fig:core+strahl_fe}
\end{figure}

To match steady state solar wind conditions, we ensure that the numerical system has zero net parallel current. Working in the rest frame of the protons, modeled as a non-drifting Maxwellian, we are left with the electron core and the strahl. Separating the electron parallel current between the two subpopulations gives
\begin{equation}
    J_\parallel = J_{\parallel,c}+J_{\parallel,s}.
\end{equation}
If the electron core with density $n_c$ is allowed to drift in the parallel direction with drift velocity $v_d$, the parallel current is 
\begin{equation}
    J_{\parallel,c} = n_c v_d.
\end{equation}
We numerically integrate Equation \ref{eqn:strahl} to obtain $J_{\parallel,s}$ and assign the electron Maxwellian core component an antiparallel (sunward) drift to exactly compensate for the strahl current and give $J_\parallel=0$:
\begin{equation}
    v_d = -J_{\parallel,s} / n_c.
\end{equation}

In Figure \ref{fig:core+strahl_fe}, we show the total electron velocity distribution at a distance of 1 AU. Several key features of this core-strahl model deviate from bi-Maxwellian representations. At a given distance, the strahl width decreases as a function of electron energy \citep[see Eq.~(18) in][]{boldyrev19}, resulting in the high energy tail narrowing with increasing parallel velocities. As distance is varied the strahl narrows from the sun out to about 1 AU due to magnetic focusing effects. At larger radial distances the strahl width saturates as diffusive effects of Coulomb collisions balance the focusing effect of decreasing magnetic field strength.

\begin{figure}
    \includegraphics[width=\columnwidth]{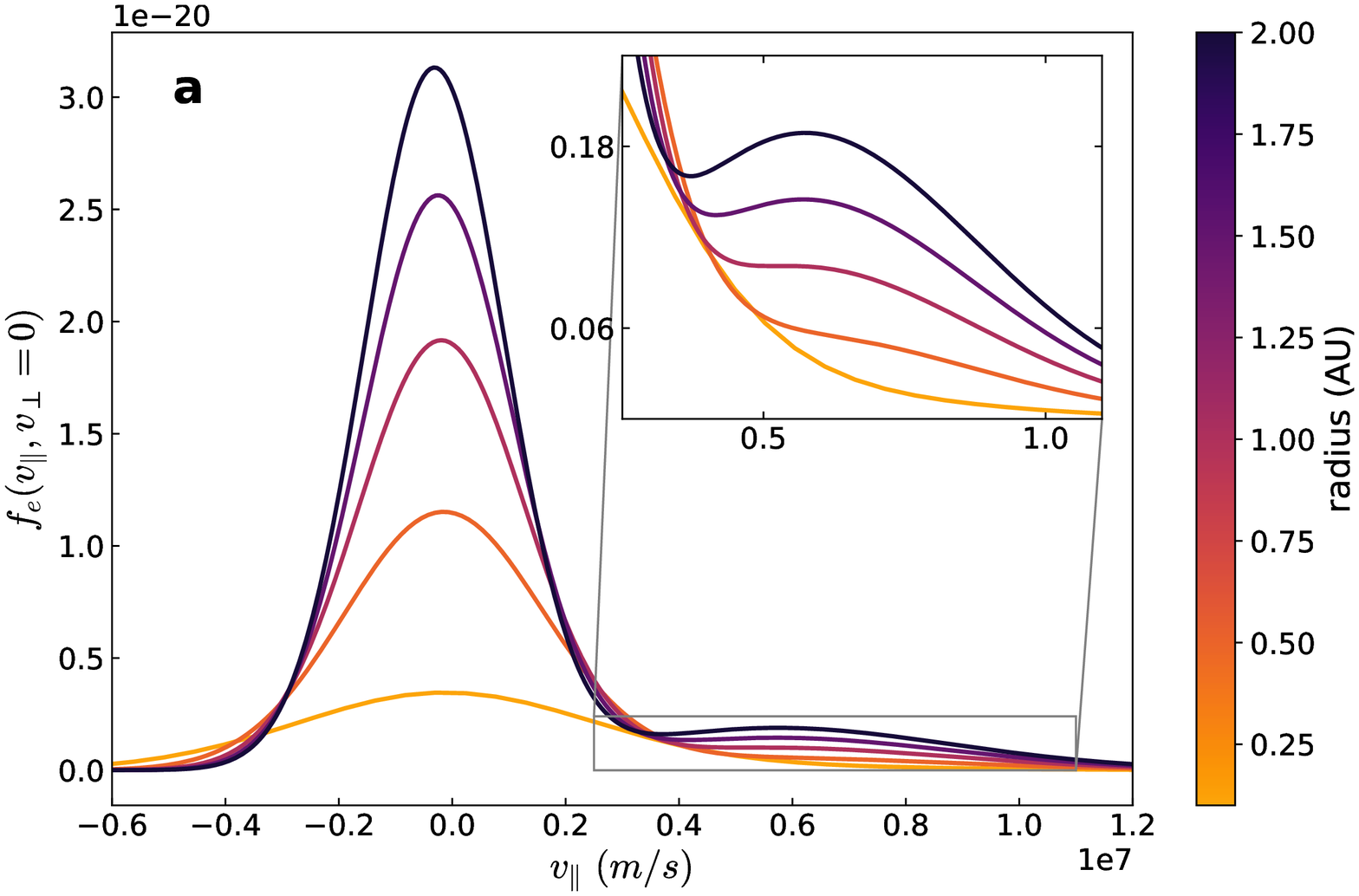}
    \includegraphics[width=\columnwidth]{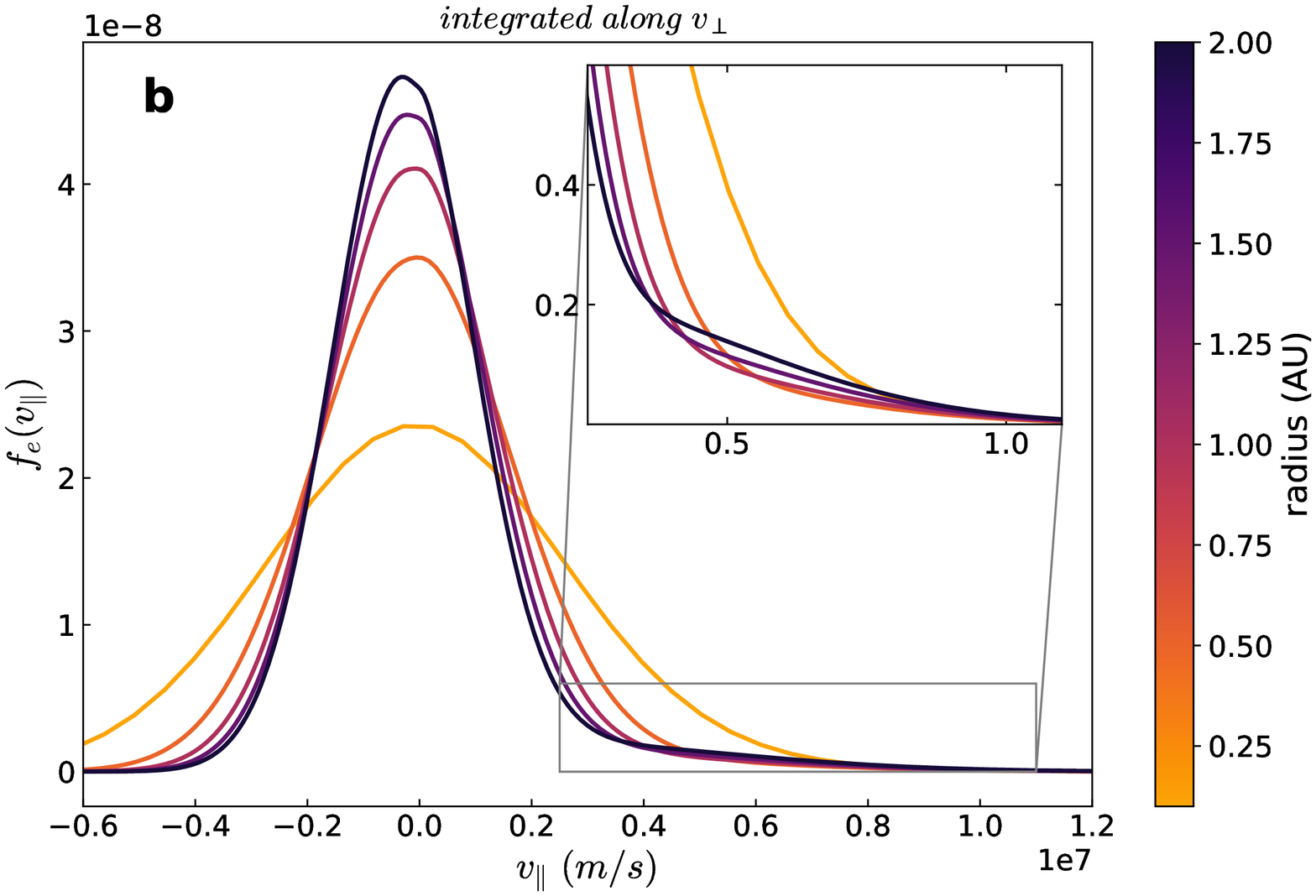}
    \caption{Radial evolution of the electron distribution function for velocities parallel to the Parker spiral shaped magnetic field. The distributions shown here are normalized to unity. As distance increases, the electron core temperature decreases with a $-1/2$ power. Simultaneously the strahl distribution increases in parallel intensity and requires a higher velocity shift between electron and ion core distributions in order to maintain net zero current. \textbf{(a)} In a cut along $v_\perp=0$ these effects compound to reveal a positive velocity gradient $\partial f / \partial v_\|$ around $5\times 10^6$~m/s. \textbf{(b)} When integrated along $v_\perp$, however, the distribution shows no positive gradients.}
    \label{fig:radial_fe_evolution}
\end{figure}

In Figure \ref{fig:radial_fe_evolution}, we show the radial evolution of the parallel electron velocity distribution function. The electron distribution obtained in the kinetic model, $f(v_\perp, v_\|; r)$,  is, in general,  not a monotonically declining function of $v_\|$. Rather, starting from a certain heliospheric distance, $f(0, v_\|; r)$ progressively develops a slight ``bump on tail'' that becomes more pronounced as the heliosperical distance increases and the core-electron temperature declines. Closer inspection of the distribution, however, reveals that the $v_\perp$-width of the strahl is a {\em decreasing} function of $v_\|$. As a result, when integrated over $v_\perp$, the distribution of the quasi one-dimensional electron beam, $\int f(v_\perp, v_\|; r)dv_\perp$, turns out to be a {\em declining} function of $v_\|$ at all distances $r$, thus not leading to electrostatic electron-beam instabilities.

\section{Stability analysis}

We perform our stability analysis using the Linear Electromagnetic Oscillations in Plasma with Arbitrary Rotationally-symmetric Distributions code (LEOPARD), which directly integrates the fully kinetic dielectric tensor components to solve for the complex frequencies of electromagnetic plasma waves \citep[]{leopard}{}. The ability to input any gyrotropic distribution function makes this code an ideal tool to study stability properties of the distribution presented in Section \ref{sect:2}. 

LEOPARD employs an iterative scheme to find the complex solution $\omega(\mathbf{k})$. The code iterates through $|\mathbf{k}|$ at fixed angle $\theta$, given a sign convention for real frequencies that $\Re[\omega]>0$ for waves propagating parallel to magnetic field lines and $\Re[\omega]<0$ for waves propagating antiparallel to the magnetic field. The imaginary part of the solution describes the stability of a given wave mode. A wave is stable when $\Im[\omega]<0$ and unstable when $\Im[\omega]>0$. To find appropriate initial guesses for the fast magnetosonic and shear Alfv\'en wave branch we refer to results in a previous stability study of electron strahl by \citet{horaites18b}.

\begin{figure}
    \includegraphics[width=\columnwidth]{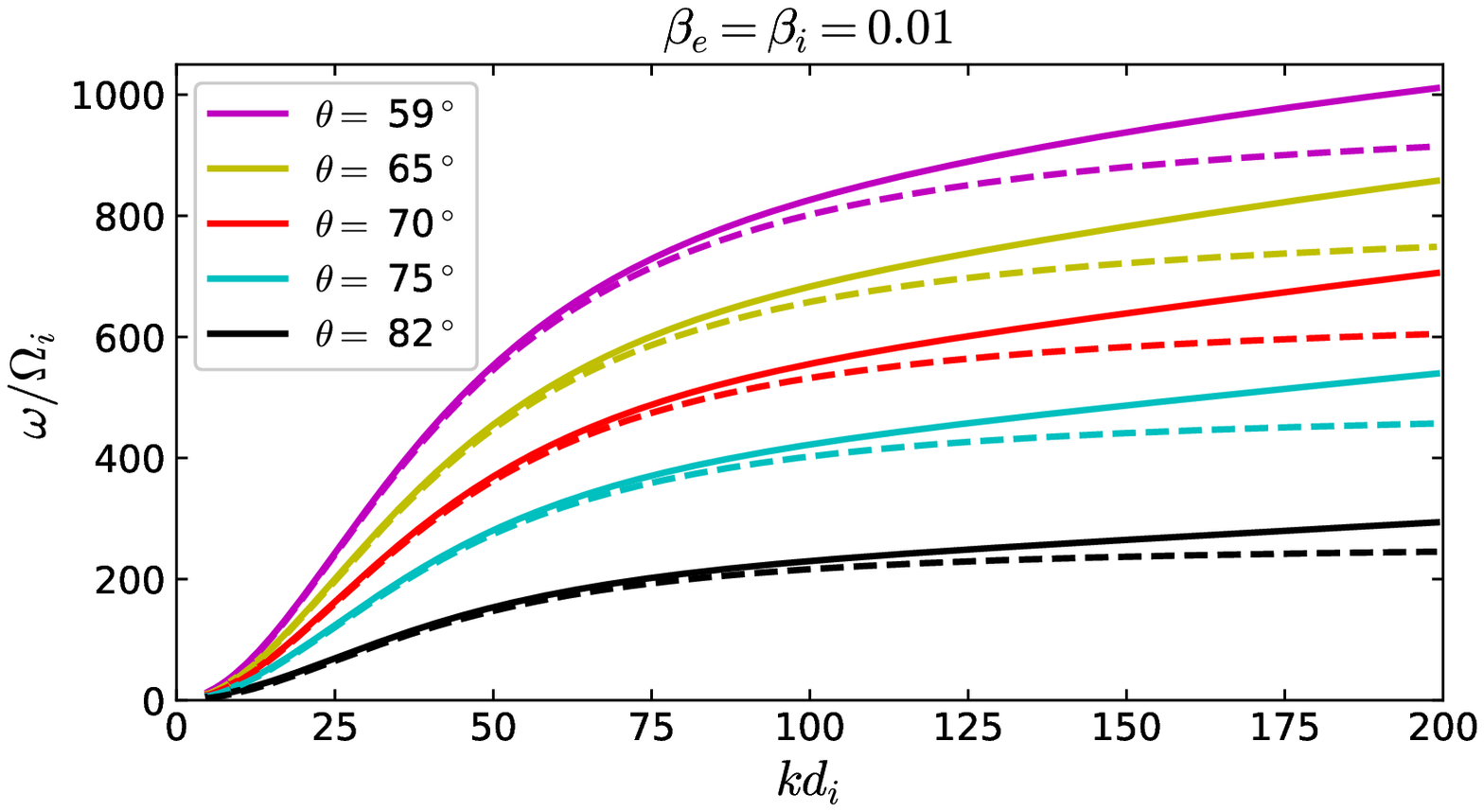}
     \includegraphics[width=\columnwidth]{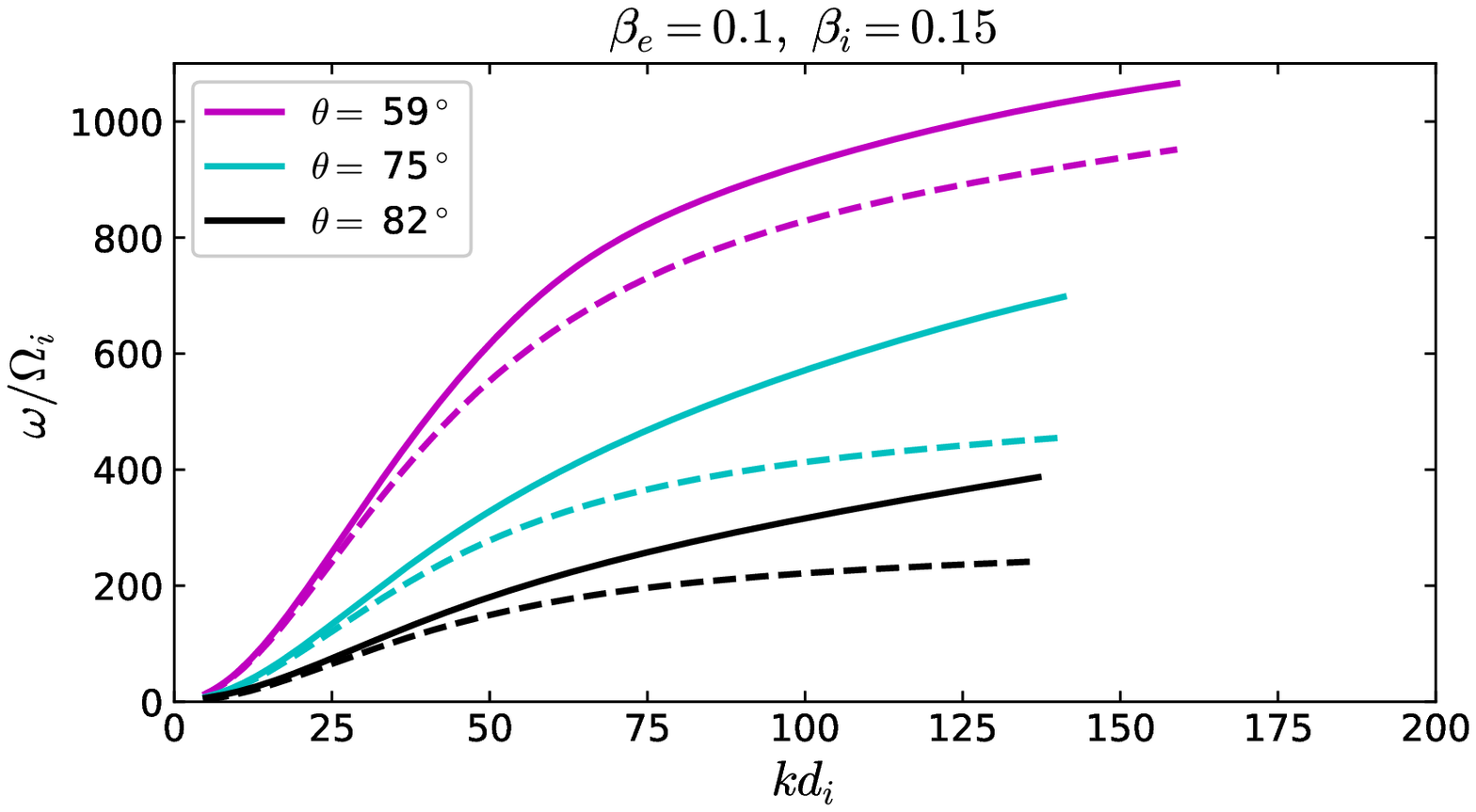}
    \caption{Comparison of analytic and numerical solutions for the whistler dispersion relations in isotropic Maxwellian plasma. Here solid lines are the numerical LEOPARD result, while the dashed lines are an analytical dispersion relation given by Equation~(\ref{eqn:analytical_disp}). For reference, the electron gyroscale is $\rho_e = \sqrt{\beta_e m_e/m_i} d_i$, which for $\theta=75^\circ$  gives $k_\perp \rho_e =1$ at $kd_i \sim 445$ for $\beta_e = 0.01$, and $kd_i \sim 140$ for $\beta_e = 0.1$. Because dissipation becomes very strong beyond this scale, curves in the lower panel are only plotted out to the $k_\perp \rho_e =1$. Note that the electron beta should be rather small for the analytic theory to provide a good approximation for the true dispersion relations.}
    \label{fig:whistler_analytical}
\end{figure}

Before we proceed with a detailed stability analysis, we would like to illustrate the advantage of the numerical solution over its analytic counterpart at small kinetic scales, which becomes essential already in the case of isotropic Maxwellian ion and electron distributions. Figure~\ref{fig:whistler_analytical} shows LEOPARD results for the kinetic-scale whistler modes, $kd_i\gg 1$. The numerical results are compared to an analytical whistler dispersion relation that takes into account the electron inertial terms \cite[e.g.,][]{Biskamp1999,Chen_2017}, and finite gyroradius corrections \citep[][]{passot_sulem_tassi_2017,passot18}: 
\begin{equation}
  \omega=\dfrac{k_\parallel k \left(1+\dfrac{\beta_e m_e}{4 m_i} k_\perp^2 \right)}{ \left(1+\dfrac{m_e}{m_i}k^2  + \dfrac{\beta_e m_e^2}{2 m_i^2}k_\perp^4   \right)^{1/2}  \Bigg( 1+\dfrac{m_e}{m_i}k^2  \Bigg)^{1/2} }.
  \label{eqn:analytical_disp}
\end{equation}
Here $\omega$ is normalized to the ion gyrofrequency $\Omega_i=eB/m_i$ and wavenumbers are normalized to the inverse ion inertial scale $d_i^{-1}=\Omega_i/v_A$. Terms $(m_e/m_i) k^2$ then correspond to electron inertial effects, while terms proportional to $\beta_e$ are finite Larmor radius corrections (it is assumed that the Larmor radius corrections are small). This model is valid for low electron beta systems, for spatial scales between the ion and electron gyroradii. For the low beta case with $\beta_e = 10^{-2}$, we find good agreement between numerical results and analytical theory as spatial scales approach the electron gyroscale. However, when we increase plasma beta to values more relevant to solar wind conditions in the inner heliosphere, $\beta_e\sim 0.1$, numerical results deviate significantly from analytical theory at small scales, especially for highly oblique angles of propagation.  These whistler results for simple isotropic Maxwellian plasmas illustrate the importance of using more precise numerical solutions when good accuracy is required. 

We now turn to the analysis of the electron core-strahl model described above. We use LEOPARD to solve for complex frequencies. We start our numerical solution at small wavenumbers, where the dissipation is weak, and continue it iteratively for larger $k$. We analyze the cases of sunward (anti-parallel to the strahl) and antisunward (parallel to the strahl) propagating waves separately.

{\em Sunward propagating waves.} We search for instabilities related to the fast magnetosonic (FM)--whistler and the shear Alfv\'en--kinetic Alfv\'en (KAW) branches.   For all these modes and for a broad range of angles ($0-89^\circ$), we observed rapidly increasing dissipation at spatial scales $kd_i\gtrsim 8$ and thus we chose to cut off iterations at this scale. For both branches, we, however, observed instabilities at $k$ values ranging from $kd_i=0.05-8$. Since the electron distribution varies with the heliospheric distance, the instability thresholds depend on the distance as well. All such critical distances turn out to be on the order of one astronomical unit. 

\begin{figure}
    \includegraphics[width=\columnwidth]{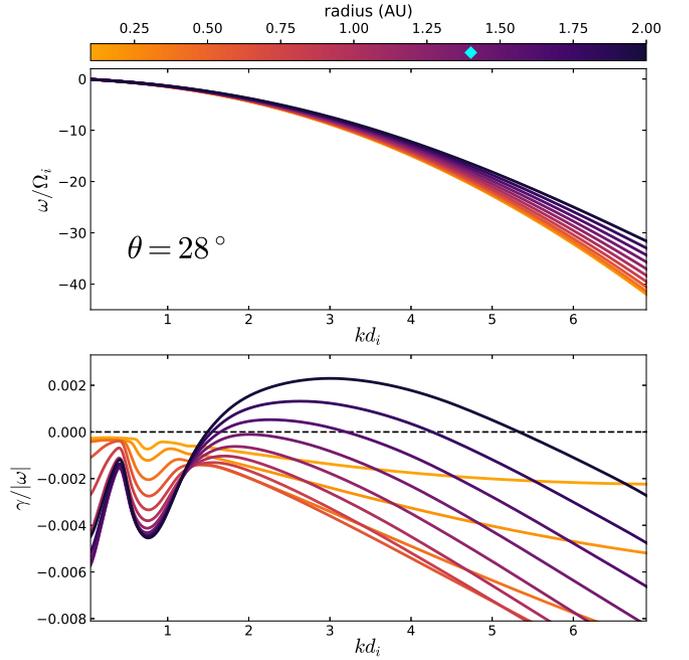}
    \caption{Radial evolution of the whistler branch at a quasi-parallel angle of $28^\circ$. This angle corresponds to the closest heliospheric distance at which the whistler instability appears for the chosen solar wind parameters. The radial distance for the instability onset is marked with a cyan diamond in the colorbar. }
    \label{fig:radial_FMwhistler_theta28}
\end{figure}

\begin{figure}
    \includegraphics[width=\columnwidth]{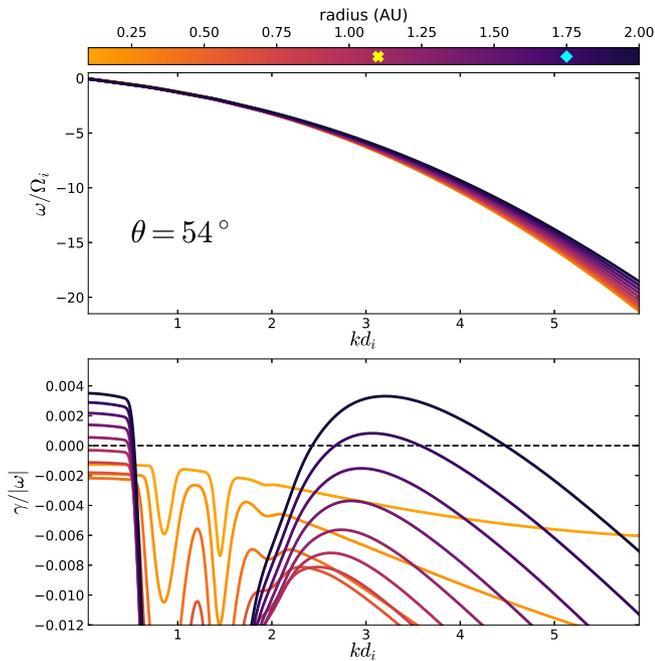}
    \caption{Radial evolution of the fast magnetosonic / whistler branch at a oblique angle of $54^\circ$. The radial onset of fast magnetosonic instability is marked with a yellow $\mathbf{x}$, and whistler instability is marked with a cyan diamond in the colorbar. This angle shows the closest distance of fast magnetosonic instability for the chosen parameters. Whistler instability begins further from the sun than at lower angles, but quickly exceeds lower angles in relative growth rate.}
    \label{fig:radial_FMwhistler_theta54}
\end{figure}

Figure \ref{fig:radial_FMwhistler_theta28} shows frequencies, growth rates, and propagation angle for the whistler mode ($kd_i > 1$) that becomes unstable at the shortest heliospheric distance. This instability was not detected in the previous analysis by \citet{horaites18b}, possibly because that work was limited to a heliospheric distance around 1 AU. At the furthest distance considered in our analysis, whistler waves are unstable in a broad range of angles from nearly parallel up to well beyond $54^\circ$ (shown in Figure \ref{fig:radial_FMwhistler_theta54}). We have also verified (but do not present here) that these low frequency anti-parallel whistlers remain unstable under slight variations in  parameters  (both  larger  and  smaller  values  of  density, temperature, mean free path at solar corona, and radial core temperature scaling), which only slightly affect quantitative results on radial instability onset. We conclude that this whistler mode is a robust instability feature for this electron distribution function.

In order to illustrate the physical mechanism of the instability, consider the whistler waves propagating at an angle $\theta=28^\circ$. In this case, the instability begins at $r\approx 1.45$AU, with a peak in growth rates at $kd_i\approx-2$ and $\omega\approx 3.9\Omega_i$. The electron core drift (with respect to the ion core, with negative sign indicating sunward drift) at this radius is $v_d\approx -3.3 v_A$. Using these values we estimate 
$k_\parallel v_d \approx 5.8 \Omega_i$ and $\omega \approx 3.9\Omega_i$. We therefore see that the instability parameters for the observed whistler instability most closely satisfy a general Landau-Cherenkov resonance condition caused by drifting beams of charged particles (electron and ion cores) given by

\begin{equation}
    \omega = \mathbf{k} \cdot \mathbf{v} \approx k_\parallel v_d.
\label{eqn:landau_resonance}
\end{equation}
(The use of $v_d$ in evaluating the resonance condition approximates a resonant velocity on the peak of the electron core. The exact resonance location occurs in a region of positive velocity gradient between the ion and electron core peaks, which explains why $k_\parallel v_d$ is a slight overestimate of the resonant value.) While we show values only for the case of $\theta=28^\circ$ propagation (which corresponds to the shortest distance at which instability occurs), the same analysis holds for other angles. The whistler modes are, therefore, excited due to a Landau-Cherenkov resonance with the drifting electron core, and not due to a cyclotron resonance with the highly energetic strahl.

\begin{figure}
    \includegraphics[width=\columnwidth]{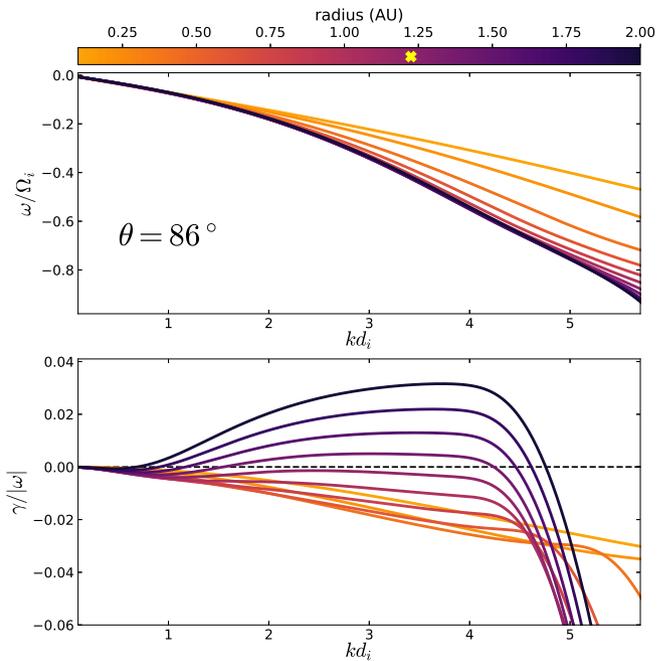}
    \caption{Radial evolution of the shear Alv\'en / KAW branch for a highly oblique angle of $86^\circ$. The radial onset of Alfv\'en instability is marked with a yellow $\mathbf{x}$. This angle shows the closest distance of KAW instability for the chosen parameters.}
    \label{fig:radial_Alfven}
\end{figure}

In the sunward propagation domain, we have also investigated other wave instabilities previously found by \citet{horaites18b}. Fast magnetosonic modes occur at oblique angles, peaking at an angle of $54^\circ$ shown in Figure \ref{fig:radial_FMwhistler_theta54}. Kinetic Alfv\'en waves are seen in a range of nearly perpendicular angles with the fastest growing mode around $86^\circ$, shown in Figure \ref{fig:radial_Alfven}. The radial onsets of the FM and KAW instabilities occur at a similar distance closer than the onset of whistler waves. Similarly to the whistler case, the resonance conditions indicate the presence of a core drift resonance, rather than a resonance with the strahl particles. Indeed, for KAWs propagating at an angle $\theta=86^\circ$, instability begins at $r\approx 1.25$AU where the core drift is $v_d\approx -2.75 v_A$, and a peak in growth rates occurs at scale $kd_i\approx-2.5$ where $\omega\approx 0.25\Omega_i$. Using these values, we estimate $k_\parallel v_d \approx 0.47 \Omega_i$, which gives close agreement to the resonance condition given by Eq.~(\ref{eqn:landau_resonance}).  For the magnetosonic waves propagating at an angle $\theta=54^\circ$, the instability begins at $r\approx 1.1$AU where $v_d\approx -2.5 v_A$, with the largest growth rate at scale $kd_i\approx-0.5$ where $\omega\approx 0.55\Omega_i$. We therefore estimate  $k_\parallel v_d \approx 0.735 \Omega_i$, which is once again quite close to the Landau resonance condition~(\ref{eqn:landau_resonance}). The obtained frequencies and wave numbers of growing modes are inconsistent with a cyclotron resonance with the strahl particles.

{\em Anti-sunward propagating waves.} We have searched for instabilities of anti-sunward propagating Alfv\'en, kinetic Alfv\'en, magnetosonic, and whistler modes in the same range of angles ($\theta= 0-89^\circ$) as for the sunward waves and did not detected any instabilities at either small ($kd_i< 1$) or large  ($kd_i>1$) wavevectors. We specifically studied the whistler waves at an angular resolution of $5^\circ$ between $5^\circ$ and $85^\circ$ for scales approaching the gyroradius (up to $kd_i\sim 60$ for oblique angles) and found that their dissipation rapidly increases at such scales. In particular, we have not detected the presence of a strahl-driven fan instability \cite[e.g.,][]{kadomtsev68,parail78,vasko_2019}, which would exist at scales $1/d_e\ll k_\perp \ll 1/\rho_e $ and wave propagation angles $\sqrt{m_e/m_i}\ll (\pi/2)-\theta \ll 1$, where the oblique whistler mode transforms, according to Eq.~(\ref{eqn:analytical_disp}), into the mode $\omega=\Omega_ek_\|/k_\perp$ (here we are using dimensional variables). In our case, such modes are dissipated very strongly, since our electron beta parameters are not very small, $\beta_e\sim 0.1\dots 0.3$, and as a consequence, the separation between the electron inertial scale $d_e$ and the gyroscale $\rho_e$ where the dissipation becomes strong, is not sufficiently large. 

\section{Conclusions}

To summarize our results, we have conducted linear stability analysis of a physically realistic distribution function for solar wind electrons. This function follows from a first principles kinetic approach and consists of a Maxwellian core and an anti-sunward directed strahl obtained as a solution of a weakly collisional drift-kinetic equation. We have focused on fast magnetosonic, whistler, Alfv\'en, and kinetic Alf\'en wave modes.  We obtained the thresholds of wave instabilities as functions of heliospheric distances and propagation angles. As an important qualitative result (that extends previous more limited analysis by \citet{horaites18b}), we have found no instabilities driven by wave resonances with the strahl particles at any distance. This suggests that a realistic electron strahl distribution obtained as a solution of a kinetic exospheric model \cite[][]{boldyrev19} (see also models developed in, e. g., \citet{landi12,bercic21}) is inherently stable and, therefore, physically realizable.

The model we consider is of course idealized in that it assumes laminar background plasma flow and ignores preexisting perturbations in magnetic flux tubes and plasma turbulence, which certainly are important effects in solar wind evolution \citep[see for example][]{Halekas_2020,Maksimovic_2020}. In this respect the considered model may not describe the shape of the strahl seen in observations when strong pitch angle scattering by background turbulence is present (such anomalous scattering effects may in principle be included in such models, see \citet{tang2018,tang20,boldyrev19}). However we believe that exospheric models are good starting points for understanding the physics of electron strahl formation. Moreover, they may shed light on the origin of kinetic instabilities and resulting kinetic-scale turbulence in the solar wind.

The instabilities that we have detected in the considered exospheric model are not strahl-resonating, rather, they are related to the relative drifts between the electron and ion cores. We found that depending on the heliospheric distance, the low-frequency kinetic-scale ($kd_i\sim 1$) kinetic Alfv\'en, magnetosonic, or whistler modes become linearly unstable. While such modes do not directly interact with the strahl particles, after an initial growth phase, their nonlinear interactions may lead to turbulent cascades, so that high-frequency whistler modes are eventually generated that are able to scatter energetic electrons, and, possibly, broaden or significantly diffuse the strahl. The topic of driven turbulent cascades of whistler modes requires nonlinear analysis that is beyond the scope of our consideration \cite[e.g.,][]{livshitz_t1972,boldyrev_95,Biskamp1999,galtier_b03}. It is interesting, however, that all the obtained instability thresholds correspond to distances on the order of $1$~AU, suggesting that kinetic-scale whistler turbulence should be effectively generated by such mechanisms only at relatively large heliospheric distances. It is also worth pointing out that the generated fluctuations may cover a broad range of propagation angles, thus suggesting that kinetic-scale turbulence is not necessarily strongly oblique or limited to directions aligned with magnetic-field lines.

\section*{Acknowledgements}
This work is supported by NSF under Grants PHY-1707272 and PHY-2010098, by NASA under Grant NASA 80NSSC18K0646, and by the Wisconsin Plasma Physics Laboratory (US Department of Energy Grant DE-SC0018266).

\section*{Data Availability}
The LEOPARD code is available from the public repository \url{https://github.com/pastfalk/LEOPARD}.
The input files used to produce the data shown in this paper is publicly available from \url{https://github.com/schroeder24/SWeVDF_stability_inputs}



\bibliographystyle{mnras}
\bibliography{mybib} 

\bsp	
\label{lastpage}
\end{document}